\documentclass[oneside,english]{amsart}

\usepackage{amsmath}
\usepackage{amssymb}
\usepackage{graphicx}
\usepackage{amsthm}


\numberwithin{equation}{section}
\numberwithin{figure}{section}
\theoremstyle{plain}
\newtheorem{thm}{\protect\theoremname}
  \theoremstyle{definition}
  \newtheorem{defn}[thm]{\protect\definitionname}

\makeindex
\makeatother

\usepackage{babel}
  \providecommand{\definitionname}{Definition}
\providecommand{\theoremname}{Theorem}

\begin{document}

%\title{Some Examples of Contextuality in Physics: Implications to Quantum
%Cognition}

\title[Contextuality in Physics and Quantum Cognition]{Some Examples of Contextuality in Physics: Implications to Quantum Cognition}

\author{J. Acacio de Barros}
\address{School of Humanities and Liberal Studies\\
San Francisco State University\\
1600 Holloway Ave. \\
San Francisco, CA 94132}
\author[J. A. de Barros, G. Oas]{Gary Oas} \address{Stanford Pre-Collegiate Studies\\
Stanford University\\
Ventura Hall\\
Stanford, CA 94305-4115}

%\author[F. Author]{First Author}
%\author[de Barros, Oas]{J. Acacio de Barros$^*$ and Gary Oas$^\dagger$}
%
%\address{$^*$School of Humanities and Liberal Studies\\
%San Francisco State University\\
%1600 Holloway Ave. \\
%San Francisco, CA 94132\\
%$^\dagger$Stanford Pre-Collegiate Studies\\
%Stanford University\\
%Ventura Hall\\
%Stanford, CA 94305-4115}

\begin{abstract}
Contextuality, the impossibility of assigning a single random variable to represent the outcomes of the same measurement procedure under different experimental conditions, is a central aspect of quantum mechanics. Thus defined, it appears in well-known cases in quantum mechanics, such as the double-slit experiment, the Bell-EPR experiment, and the Kochen-Specker theorem. Here we examine contextuality in such cases, and discuss how each of them bring different conceptual issues when applied to quantum cognition. We then focus on the shortcomings of using quantum probabilities to describe social systems, and explain how negative quasi-probability distributions may address such limitations. 
\end{abstract}

\maketitle

\section{Introduction}

Contextuality is defined as the impossibility of assigning a single random variable to represent the outcomes of a measurement under different
experimental conditions (thought of as contexts)~\cite{spekkens_contextuality_2005,larsson_necessary_1998,dzhafarov_contextuality_2014-1,dzhafarov_contextuality_2014,de_barros_unifying_2014}.
More precisely, say you have a series of experimental conditions where you measure a property $P$, whose outcome (out of several runs) is represented by the random variable $\mathbf{P}$. Imagine that, for
one of those conditions, $P$ is measured together with other properties
$A_{1},A_{2},\ldots,A_{n}$ (whose outcomes are also represented by
random variables $\mathbf{A}_{1},\mathbf{A}_{2},\ldots,\mathbf{A}_{n}$),
but also imagine that, for another condition, $P$ is measured with
$B_{1},B_{2},\ldots,B_{n'}$, and finally assume that it is not possible
to create an experiment where all properties $P,A_{1},A_{2},\ldots,A_{n},B_{1},B_{2},\ldots,B_{n'}$
are measured simultaneously. Contextuality, as defined informally
above, is thus the impossibility of finding a probability space $\left(\Omega,\mathcal{F},p\right)$
for $\mathbf{P},\mathbf{A}_{1},\ldots,\mathbf{A}_{n},\mathbf{B}_{1},\ldots,\mathbf{B}_{n'}$
compatible with the distributions observed experimentally. 

As an example, take a simple situation where we have three properties
$X$, $Y$, and $Z$ corresponding to true or false statements. Observing
such properties is modeled by $\pm1$-valued random variables, $\mathbf{X}$,
$\mathbf{Y}$, and $\mathbf{Z}$. Assume now that we only observe
two properties at a time, but never all three together. Assume additionally
that they are seen as perfectly anti-correlated to each other for
each experimental condition, i.e. 
\begin{equation}
E\left(\mathbf{XY}\right)=E\left(\mathbf{XZ}\right)=E\left(\mathbf{YZ}\right)=-1.\label{eq:xyz-maximal}
\end{equation}
Clearly no probability space giving those correlations exists, since
a $\omega\in\Omega$ giving $\mathbf{X}\left(\omega\right)=1$ implies,
from the first and second expectations in (\ref{eq:xyz-maximal}),
$\mathbf{Y}\left(\omega\right)=-1$ and $\mathbf{Z}\left(\omega\right)=-1$,
which contradicts the third expectation. However, if we relabel the
variables, in the spirit of References \cite{peres_quantum_1995,dzhafarov_random_2013,dzhafarov_all-possible-couplings_2013,dzhafarov_qualified_2014,dzhafarov_contextuality_2014-1},
making the fact that they were measured in a pairwise way explicit,
it is possible to construct a $\left(\Omega,\mathcal{F},p\right)$.
For instance, if we have a new set of random variables $\mathbf{X}_{\mathbf{Y}}$,
$\mathbf{X}_{\mathbf{Z}}$, $\mathbf{Y}_{\mathbf{X}}$,\textbf{ $\mathbf{Y}_{\mathbf{Z}}$},
$\mathbf{Z}_{\mathbf{X}}$, and $\mathbf{Z}_{\mathbf{Y}}$, such that
correlations in (\ref{eq:xyz-maximal}) are now 
\begin{equation}
E\left(\mathbf{X_{\mathbf{Y}}Y_{X}}\right)=E\left(\mathbf{X_{\mathbf{Z}}Z_{\mathbf{X}}}\right)=E\left(\mathbf{Y_{\mathbf{Z}}Z_{\mathbf{Y}}}\right)=-1,\label{eq:xyz-maximal-CdB}
\end{equation}
the sampling of an $\omega\in\Omega$ leads to no contradictions. It
is straightforward that the contradiction from (\ref{eq:xyz-maximal})
comes from assuming that the value of, say, $\mathbf{X}$ when measured
with $\mathbf{Y}$ is the same as its value when measured with $\mathbf{Z}$,
i.e., it does not depend on the experimental context provided by the
simultaneous measurement of the pairs. 

Contextuality as defined here is ubiquitous in quantum mechanics,
and may be at the heart of what defines quantum systems, as opposed
to classical ones (see \cite{oas_survey_2015} in this volume). Examples of
contextual quantum systems, some of them discussed in more details
in Section \ref{sec:Contextuality-in-QM}, are successive measurements
of spin \cite{gerlach_experimentelle_1922,peres_quantum_1995}, the
double-slit experiment \cite{dirac_principles_1947}, the Leggett-Garg
experiment \cite{leggett_quantum_1985}, the EPR-Bell experiment \cite{einstein_can_1935,bohm_quantum_2012},
and the Kochen-Specker system of observables \cite{kochen_problem_1967}.
So, it should come as no surprise that the mathematical formalism
developed to describe quantum systems is suitable to describe (at
least certain) contextual systems. 

It is perhaps for this reason that such formalism was successfully
applied to social systems with a certain degree of success \cite{busemeyer_quantum_2012,haven_quantum_2013}.
Social systems, because of their contextuality, lack a joint probability
distribution, and the contextual calculus of probability of quantum
mechanics seems to offer a suitable framework for situations in which
standard probability theory fails. For example, Savage's famous Sure-Thing-Principle
(STP) \cite{savage_foundations_1972}, a consequence of classical
probability theory, are violated by human decision-makers \cite{tversky_disjunction_1992,shafir_thinking_1992}.
Khrennikov and Haven \cite{khrennikov_quantum_2009} applied principles
of quantum interference and showed that certain superpositions of
quantum-like states representing mental processes could be used to
describe the experimentally verified violation of STP. 

So, it is possible the relationship between quantum mechanics, its formalism,
and social phenomena goes beyond a simple analogy, but points to a
deeper relationship between the Hilbert space formalism and the description
of determinate contextual systems. However, arguments exist that certain
social processes may not be describable by the quantum formalism \cite{de_barros_joint_2012,de_barros_decision_2014,de_barros_beyond_2015},
but instead by other contextual probability theories. Be that as it
may, given the increasing importance of the quantum formalism in the
social sciences \cite{khrennikov_ubiquitous_2010,busemeyer_quantum_2012,haven_quantum_2013},
we believe that a distinction of the different quantum process that
exhibit contextuality should be fruitful. Here we analyze three different
quantum systems that are contextual, and show that each have contextuality
that present different conceptual features. 

This paper is organized the following way. First, in Section \ref{sec:Contextuality-in-QM},
we examine contextuality in quantum mechanics, starting with the famous
double-slit experiment, and then moving to the Bell-EPR experiment
and the Kochen-Specker theorem. In Section \ref{sec:Contextuality-in-QC},
we discuss contextuality in quantum cognition, and how it relates
to the examples discussed from physics. Finally, in Section \ref{sec:Describing-QC}
we present a particular alternative model of extended probabilities
that is suitable for some of the situations discussed in quantum cognition.

\section{Contextuality in QM\label{sec:Contextuality-in-QM}}

A fundamental question in physics is what makes quantum systems different
from classical ones (see our other contribution to this volume \cite{oas_survey_2015}).
At the bottom of it seems to be the apparent impossibility of describing
quantum systems in terms of concepts from classical physics, such
as particles and fields. 

In the early days of the quantum revolution, physicists attempted
to describe the microscopic phenomena observed in terms of what we
know nowadays as classical ideas. But soon many realized that causality,
one of the main tenets of classical physics, was not compatible with
the quantum world. For instance, Rutherford's radioactive decay formula
was seen as corresponding to a memoryless Poissonian process, and
that therefore atoms who were about to decay had the same state as
those whose decay would happen much later. Subsequently,
shortly after Bohr published his theory explaining the spectrum of
the Hydrogen atom, Rutherford himself remarked that Bohr's theory
had a problem with causality \cite{pais_inward_1986}. Classical physics,
a causal theoretical structure with its description in terms of phase-space 
states and Hamiltonian dynamics, was thought to not be able
to account for what were essentially probabilistic processes. As such,
the unavoidable probabilistic character of quantum mechanics became
a topic of intense discussion and research. 

Connected to this discussion was the idea that quantum probabilities
could have their origins in the impossibility of simultaneously observing
two complementary quantities, such as momentum and position of a particle.
For instance, for Heisenberg, a measurement of position would cause
a random disturbance on the momentum in such a way that knowing the
position of a particle at a time $t_{0}$ would make a prediction
of its position at time $t>t_{0}$ an impossible task. This perspective
evolved into a viewpoint some physicists put forth that two incompatible
properties could not \emph{exist }at the same time, brought about
by the discovery of the spatial quantization (spin). 

To see this,
imagine that spin is represented by a three-dimensional vector random
variable $\mathbf{\mu}\left(\omega\right)$ (here we follow \cite{peres_quantum_1995}).
If a measurement of spin, say, in the direction $\hat{\mathbf{z}}$,
simply reveals the value of such random variable in such direction,
without disturbing it, then its result would be $\mbox{\ensuremath{\mu}}\cdot\hat{\mathbf{z}}$,
which experimentally can take only values $\pm1$ (here we use units
where $\hbar/2=1$). However, there is nothing special about the direction
$\hat{\mathbf{z}}$: the Stern-Gerlach (SG) apparatus measuring spin
could be pointing in any direction of our choice. Let us assume two
other possible measurement directions, $\hat{\mathbf{e}}_{1}$ and
$\hat{\mathbf{e}}_{2}$, such that they are each at 120 degrees from
each other, i.e.
\begin{equation}
\hat{\mathbf{z}}+\hat{\mathbf{e}}_{1}+\hat{\mathbf{e}}_{2}=0.\label{eq:three-vectors-120}
\end{equation}
Since any direction of the SG apparatus will result in quantized spin
(we assume the source is a proper mixture), we have at once that $\mbox{\ensuremath{\mu}}\cdot\hat{\mathbf{z}}$,
$\mbox{\ensuremath{\mu}}\cdot\hat{\mathbf{e}}_{1}$, and $\mbox{\ensuremath{\mu}}\cdot\hat{\mathbf{e}}_{2}$
have values $\pm1$. But 
\begin{equation}
\mathbf{\mu}\left(\omega\right)\cdot\hat{\mathbf{z}}+\mbox{\ensuremath{\mu}}\left(\omega\right)\cdot\hat{\mathbf{e}}_{1}+\mbox{\ensuremath{\mu}}\left(\omega\right)\cdot\hat{\mathbf{e}}_{2}=\mathbf{\mu}\cdot\left(\hat{\mathbf{z}}+\hat{\mathbf{e}}_{1}+\hat{\mathbf{e}}_{2}\right),\label{eq:contradiction-spin-120}
\end{equation}
and we reach an apparent contradiction, since the left hand side can
only take values $\pm3,\pm1$ but the right hand side is zero because
of (\ref{eq:three-vectors-120}). It is clear that the contradiction
comes from the assumption that the $\omega$ for each experiment (for
the three different directions) is the same, since this is what is
required to go from the left to the right hand side of (\ref{eq:contradiction-spin-120}).
Thus, if we assume that spin $\mathbf{\mu}\left(\omega\right)$ exists
before the measurement, the act of measuring it in one direction,
say $\hat{\mathbf{x}}$, changes it to a new $\mathbf{\mu}\left(\omega\right)$,
and the process of measurement does not ``reveal'' the actual state,
but instead changes it to a new state with different properties from
the original one. The relationship between $\mathbf{\mu}$ and $\omega\in\Omega$
can be thought as a hidden variable theory of the outcomes of spin,
using the terminology of physics \cite{bell_speakable_2004}, and
the dependency of $\omega$ on the experimental setup makes this theory
contextual. So, even in the simple case of having consecutive
measurements of spin, we see the impossibility of treating outcomes of experiments
as context-independent. 

\begin{figure}

\begin{centering}
\includegraphics[scale=0.7]{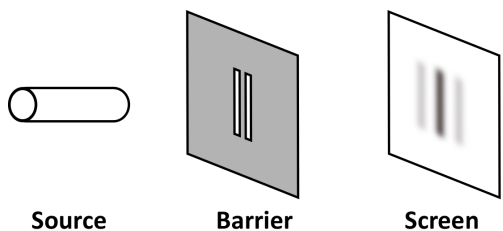}\protect\caption{\label{fig:Double-slit-experiment}Double-slit experiment. A source
on the left sends light onto a barrier with two slits cut close to
each other. An interference pattern appears on the screen to the right. }

\par\end{centering}

\end{figure}

But perhaps the best-known example of contextuality (again, in the
sense used above) in quantum mechanics is the famous double-slit experiment\index{double-slit experiment}
\cite{dirac_principles_1947}. In classical physics, the double-slit
experiment, attributed to Thomas Young \cite{born_principles_1999},
was used to demonstrate interference, thus ``falsifying'' the
corpuscular theory of Newton and supporting the wave theory of light.
In it, light impinges on a barrier where two small and parallel slits
are cut, allowing a small amount of the light to go through  and
reach a screen at the other side (see Figure \ref{fig:Double-slit-experiment}).
 Because of its wave character, an interference pattern emerges at
the screen, due to differences of phase at each screen location in
a way consistent with the geometry of the setup. This interference
pattern seems to be incompatible with the Newtonian particle theory
of light, since particles arriving at a point on the screen did
not interact with both slits, and therefore a concept of phase difference for particles 
would be meaningless. 

In its quantum mechanical version, the double slit needs to be examined
in a new light (no pun intended). As is well-known, Einstein used
the notion that light was made of particles to explain the photo-electric
effect, and evidenced mounted in the early days that light was actually
made of particles, called photons \cite{pais_inward_1986}. But if
light is made of particles, what should we conclude from Young's experiment?
An initial hypothesis was that interference was a collective effect of many
photons, similar to sound waves being a collective effect of many
atoms. However, this idea showed to be inconsistent with experiments where
the light intensity was so low that effectively only one photon at
a time was present between the slits and the screen (or photographic
plate, in this case). So, photons seem to present self-interference,
a quite mysterious property. In fact, in his famous Lectures on Physics,
Richard Feynman stated that the double-slit experiment contained the
``only mystery'' of quantum mechanics \cite{feynman_feynman_2011}. 

To see how the double-slit experiment provides contextuality, let
us examine it in a simplified version. In the double-slit experiment,
photons are detected on a screen, thus providing a continuum of locations,
with a corresponding complicated mathematical description 
(see \cite{suppes_random-walk_1994,suppes_diffraction_1994} and references therein). But what
makes the double slit mysterious to Feynman are interference effects,
and interference can be studied without resorting to such continuum,
in a setup called the Mach-Zehnder interferometer\index{Mach-Zehnder interferometer}
(MZI). So, here we analyze in more detail the contextuality of quantum
systems in the MZI. 

\begin{figure}
\begin{centering}
\includegraphics[scale=0.65]{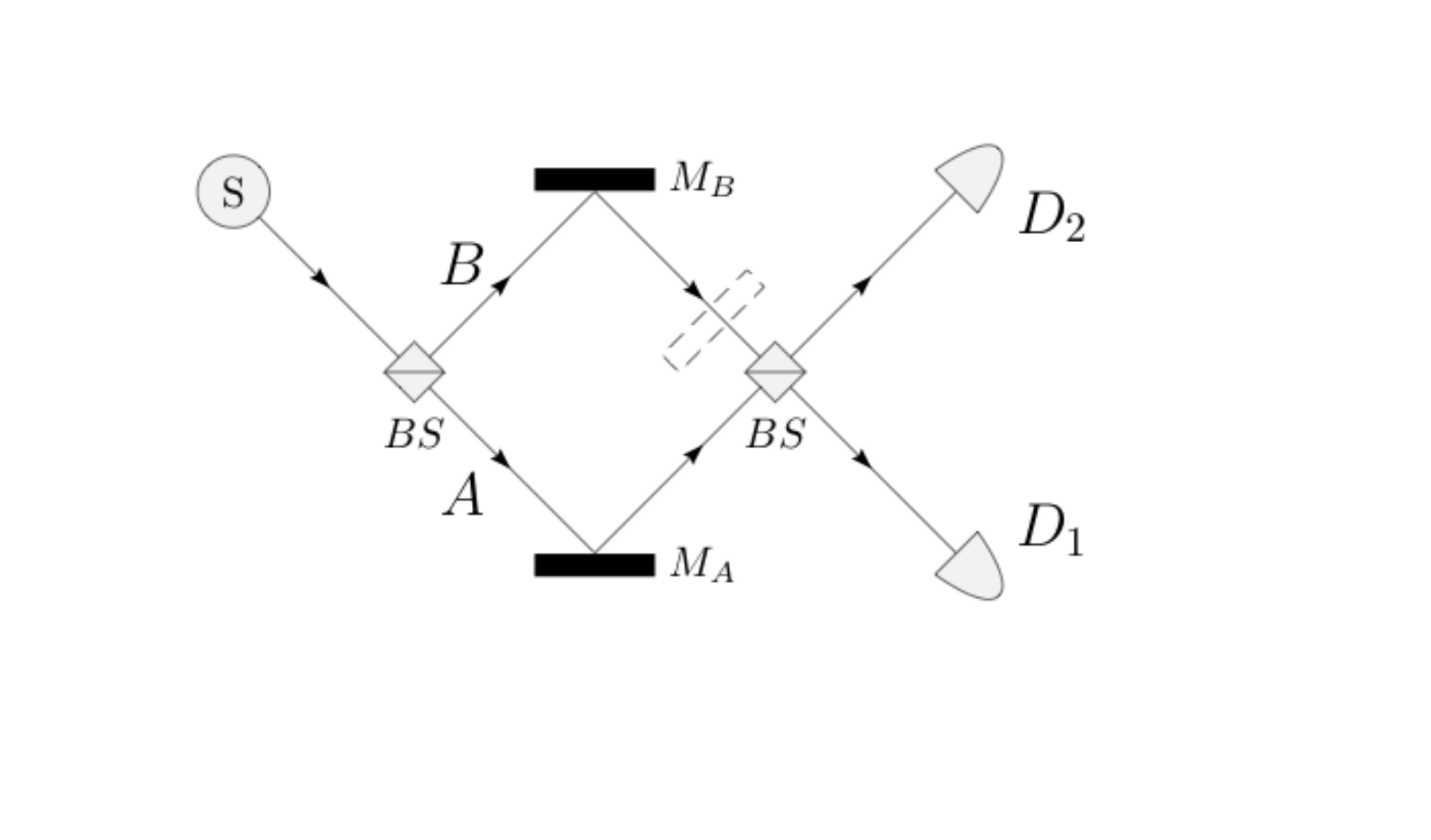}
\par\end{centering}

\protect\caption{\label{fig:Mach-Zehnder-Interferometer}Mach Zehnder Interferometer
(MZI). A light source $S$ sends a light beam that impinges on the
first beam splitter $BS$. The beam is them divided by $BS$ into
equal-intensity beams that travel to both arms (paths) $A$ and $B$
of the interferometer, reflecting on mirrors $M_{A}$ and $M_{B}$,
respectively. The beams from arms $A$ and $B$ are then recombined
in the second beam splitter. The outcomes are the two beams detected
at $D_{1}$ and $D_{2}$. }

\end{figure}

In the MZI, a light source is directed toward a beam splitter that
divides the beam into two distinct beams of equal intensity (see Figure
\ref{fig:Mach-Zehnder-Interferometer}).
 Beam splitters have the important characteristic that light beams
reflected by them are phase-shifted by $\pi/2$, whereas the beam's
phase going through is not affected, and therefore a $\pi/2$ phase
difference exists between beams in arm $A$ and $B$ of Figure \ref{fig:Mach-Zehnder-Interferometer}.
After the first beam splitter, some mirrors redirect the beams to
another beam splitter, and the beams are recombined, adding once again
a $\pi/2$ phase to the reflected beam. Mathematically, we can describe
each beam impinging on the second $BS$ with a sine function
\begin{eqnarray*}
\psi_{A} & = & \frac{A}{2}\sin\omega t,\\
\psi_{B} & = & \frac{A}{2}\sin\left(\omega t+\frac{\pi}{2}\right)=\frac{A}{2}\cos\omega t,
\end{eqnarray*}
where $A$ is the amplitude of the source $S$ and $\omega$ its frequency.
After the second beam splitter, we have 
\begin{eqnarray*}
\psi_{D_{1}} & = & \frac{A}{2}\sin\left(\omega t+\frac{\pi}{2}\right)+\frac{A}{2}\cos\omega t\\
 & = & A,\\
\psi_{D_{2}} & = & \frac{A}{2}\sin\omega t+\frac{A}{2}\cos\left(\omega t+\frac{\pi}{2}\right)\\
 & = & 0.
\end{eqnarray*}
We then see the main characteristic of the MZI: interference effects
give a zero amplitude at $D_{1}$ and amplitude $A$ at $D_{2}$.
We remark that there is an underlying assumption in our derivation
above, namely that the length of the interferometer arms $A$ and
$B$ were exactly the same. Were they any different, and the phase
relations would not match exactly,  interference would not be perfect
as above (with $0$ and $A$ amplitudes)\footnote{That is why interferometers are very useful for  measuring
distances accurately.}. 

The above description is the classical one for light waves. But what
happens if we have classical (non-quantum) particles in a similar
setup to the MZI? Imagine we send one particle at a time through the
MZI. Since no concept of phase or phase relation exists for a classical
particle, the first beam splitter would simply reflect it to $A$
or $B$, with probability $1/2$ to go to $A$ and $1/2$ to $B$.
Once in the second beam splitter, the particle (coming from\emph{
either} $A$ or $B$) would be randomly sent to either detector. Thus,
for classical particles we should expect probability $1/2$ of observing
the particle in $D_{1}$ or $D_{2}$ (but never on both detectors!). 

Now, what to make of the MZI for quantum particles? If we send a single
photon through a MZI, a photodetector placed on either arm of it will
reveal the characteristic of a particle: a click on either $A$ or
$B$, but never both. Furthermore, if a photodetector is placed on
either $A$ or $B$, the outcomes of a measurement on $D_{1}$ and
$D_{2}$ are exactly what we expect from a particle: the photon shows
up on each of those detectors with probability $1/2$. However, if
no detectors are placed on $A$ or $B$, the photon shows zero probability
of detection on $D_{2}$ and probability $1$ on $D_{1}$. In other
words, in the absence of detectors on $A$ or $B$, the photon behaves
as if it were a wave, carrying information about the relative phases
of the MZI's geometry. 

The disturbing aspects of this wave/particle duality for photons has
been discussed at length for almost one hundred years, and the interested
readers are referred to the many excellent sources (we can particularly
recommend the historical account found in Abraham Pais's volume \cite{pais_inward_1986}).
Here we focus only on the contextual aspects of it. To see them, we
start with two $\pm1$-valued random variables, $\mathbf{P}$ and
$\mathbf{D}$, representing the which-path information and detection,
respectively. $\mathbf{P}$ is defined such that $\mathbf{P}=1$ if
the particle is detected on $A$ and $-1$ otherwise, whereas $\mathbf{D}$
is defined such that $\mathbf{D}=1$ if the photon was detected in
$D_{1}$ and $-1$ if detected in $D_{2}$. 

We have for the MZI two different experimental conditions: one in
which no detector is placed on $A$ or $B$ and another where a detector
is placed on either $A$ or $B$ (or both), thus yielding which-path
information encoded in the outcomes of $\mathbf{P}$. Measuring $\mathbf{D}$
under the no-which-path condition results in $E\left(\mathbf{D}\right)=1$,
whereas measuring it together with $\mathbf{P}$ gives as marginal
expectation the result $E\left(\mathbf{D}\right)=0$. Thus, according
to the above definition of contextuality, the random variable $\mathbf{D}$
is contextual\footnote{Dzhafarov and Kujala refer to this type of strong contextuality as
\emph{direct influences}, meaning that the measurement of $\mathbf{P}$
directly influences the outcomes of random variable $\mathbf{D}$
\cite{dzhafarov_all-possible-couplings_2013,dzhafarov_qualified_2014}.
They, on the other hand, reserve the label \emph{contextual} to refer
to other cases where the influences of $\mathbf{P}$ over $\mathbf{D}$
cannot be accounted for by direct influences. }. 

The double-slit experiment  provides a dramatic type of contextuality,
but it is not as surprising as Feynman makes it seem. In fact, not
only can we ``solve'' the mystery of quantum mechanics for this
case, if we were to accept a (contextual) hidden-variable theory such as Bohm's
\cite{bohm_suggested_1952-1,bohm_suggested_1952}\footnote{Such hidden variable theory provides a mechanism that accounts for
the experimental outcomes of the double-slit experiment. As expected,
Bohm's theory cited here is a contextual theory, in the sense that
the hidden variable needs to be context-dependent. }, but we can also clearly understand the possibility of a direct influence
from $\mathbf{P}$ to $\mathbf{D}$. In fact, that is exactly what
Heisenberg tried to do with his analysis of the wave/particle duality in the
double-slit experiment by using what is now known as the Heisenberg
microscope \cite{heisenberg_physical_1949}. Thus, even though it
has at its core one of the main characteristics of QM, i.e. the interference
of particles, it is far from containing its only mystery. 

\begin{figure}
\begin{centering}
\includegraphics[scale=0.5]{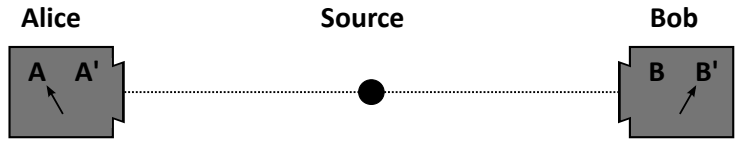}\protect\caption{\label{fig:Bell-EPR-experiment}Bell-EPR experiment. A source emits
two photons, one toward Alice's lab and another toward Bob's. Each
experimenter can make a decision on which direction of spin to measure,
represented in the figure by the settings $A$ and $A'$ for Alice
and $B$ and $B'$ for Bob. Outcomes of measurements are $\pm1$,
with equal probabilities. }

\par\end{centering}

\end{figure}

Perhaps a deeper mystery comes from a variation in the experiment
proposed by Einstein, Podolsky, and Rosen's seminal 1935 paper on
the completeness of quantum mechanics \cite{einstein_can_1935}, now
simply known as EPR\index{EPR experiment}. In Bohm's version of the
EPR argument \cite{bohm_quantum_2012}, two correlated photons\footnote{Bohm's version actually used spin $1/2$ particles, not photons, but
for our purposes we can use photons.} in the state
\begin{equation}
|\psi\rangle=\frac{1}{\sqrt{2}}\left(|++\rangle+|--\rangle\right),\label{eq:psi-EPR}
\end{equation}
where $|++\rangle$ corresponds to the state where both photons have
vertical polarization and $|--\rangle$ horizontal, are sent in two
different directions. In one direction, an experimenter, Alice, chooses
whether to measure the linear polarization of the photon or not, and if she
does, she either observes $+$ or $-$ (see Figure \ref{fig:Bell-EPR-experiment}).
 In the other direction, Bob can also measure polarization in the same direction as Alice.
However, if Alice already did, Bob does not need to, because if Alice
measures $-$ Bob knows for sure that the photon getting to him will
also be $-$ (due to the correlations contained in the state (\ref{eq:psi-EPR})). 

Up to now there is no mystery from the correlated outcomes of Alice
and Bob. What we know is that for state (\ref{eq:psi-EPR}) the outcomes
of experiments for Alice are the same as for Bob's. But that, according
to EPR, presents a problem for QM, since Alice and Bob can perform
their measurements as far away from each other as possible. If they
do, argue EPR, then it is possible to determine the value of the spin
for Bob's particle, say, without ever affecting its state, since all
we need to do is use Alice's value. Alice's measurement does not affect
the state at Bob's because special relativity forbids faster-than-light
interactions. So, continue EPR, the values of the polarization for
both particles must come from some state of the system that is not
represented in $|\psi\rangle$, and therefore the QM description of
nature is incomplete. 

A theory that completes QM, in the sense given by EPR, is called a hidden-variable theory, and it so happens that (local) hidden-variable theories are not compatible with the predictions of QM. The first person to point out an empirical incompatibility between QM and (local) hidden variable theories was John Bell. In a seminal paper \cite{bell_einstein-podolsky-rosen_1964}, Bell derived a set of inequalities that were necessary for local hidden variable theories, and proceeded to show that for certain quantum states those inequalities were violated.
More than a decade later, Aspect, Grangier, and G\'{e}rard \cite{aspect_experimental_1981}
showed, in a \emph{tour de force} experiment where for the first time
correlations between spacelike separated events were recorded, that
the quantum mechanical predictions were correct. 

To understand Bell's results, let us examine the setup he discussed.
It can be shown that a hidden-variable $\mathbf{\lambda}$ explaining
the experimental outcomes of polarization for Alice and Bob exists
if and only if a joint probability distribution exits for all possible
outcomes \cite{suppes_when_1981,suppes_collection_1996}. To show
that a hidden-variable theory is not compatible with QM, we need to
show an QM example that does not allow a joint probability distribution.
However, this setup needs to have some constraints, since we saw that
the double-slit experiment  does not have a joint probability but
is also compatible with a (contextual) hidden-variable theory. A suggestion
for this constraint is present in EPR's example: the outcomes of a
variable $\mathbf{A}$ cannot be (superluminally) influenced by what
happens at another far away location. In other words, from EPR's point
of view relativity theory is incompatible with events that are contextual
and spacelike separated. 

We now proceed to show Bell's argument and setup. Imagine that we
have now two possible experimental (and incompatible) measurements
of polarization for Alice in two different directions, and the same
for Bob (not necessarily the same directions as Alice). Let us represent
the outcomes of measurements with $\pm1$-valued random variables,
namely $\mathbf{A}$ and $\mathbf{A}'$ for both of Alice's directions,
and $\mathbf{B}$ and $\mathbf{B}'$ for Bob's (see Figure \ref{fig:Bell-EPR-experiment}). We can construct a
random variable $\mathbf{S}$ defined simply by 
\begin{equation}
\mathbf{S}=\mathbf{AB}+\mathbf{A}'\mathbf{B}+\mathbf{A}\mathbf{B}'-\mathbf{A}'\mathbf{B}'.\label{eq:S}
\end{equation}
If a joint probability exists, for each $\omega\in\Omega$ there are
associated outcomes for $\mathbf{A}$, $\mathbf{A}'$, $\mathbf{B}$,
and $\mathbf{B}'$ and a corresponding probability. Since $\mathbf{A}$,
$\mathbf{A}'$, $\mathbf{B}$, and $\mathbf{B}'$ are $\pm1$-valued
random variables, it follows that for all possible combinations of
values for $\mathbf{A}$, $\mathbf{A}'$, $\mathbf{B}$, and $\mathbf{B}'$
the value of $\mathbf{S}$ is either $2$ or $-2$ (for example, if
$\mathbf{A}=1$, $\mathbf{A}'=-1$, $\mathbf{B}=-1$, and $\mathbf{B}'=1$,
then $\mathbf{S}=-1+1+1+1=2$). Therefore, the expected value of $\mathbf{S}$
must be a number between $-2$ and $2$, i.e. 
\begin{equation}
-2\leq\left\langle \mathbf{AB}\right\rangle +\left\langle \mathbf{A}'\mathbf{B}\right\rangle +\left\langle \mathbf{A}\mathbf{B}'\right\rangle -\left\langle \mathbf{A}'\mathbf{B}'\right\rangle \leq2.\label{eq:CHSH}
\end{equation}
This inequality is one of the Clauser-Horne-Shimony-Holt (CHSH) inequalities\index{CHSH inequalities}
\cite{clauser_proposed_1969} (the others are obtained simply by moving
the ``$-$'' sign in (\ref{eq:S}) to other terms), and a joint
probability distribution exits if and only if they are satisfied \cite{fine_hidden_1982}.
It is possible to show that, for a proper choice of angles of measurement
for Alice and Bob, their observed correlations result in an $\left\langle \mathbf{S}\right\rangle =2\sqrt{2}$,
which violates (\ref{eq:CHSH}). Therefore, no joint probability distribution
exists, and as consequence, no hidden-variable theory exists that
explains the correlations between the observables. Furthermore, because
no joint exists, the assumption that an $\mathbf{A}$ under experimental
condition where it is measured together with $\mathbf{B}$ (as in
$\left\langle \mathbf{AB}\right\rangle $) is the same as when it
is measured with $\mathbf{B}'$ is not correct: the system $\mathbf{A}$,
$\mathbf{A}'$, $\mathbf{B}$, and $\mathbf{B}'$ is contextual. 

It is hard to overplay the importance of Bell's results. For example,
Henry Stapp famously stated it to be ``the most profound discovery
of science'' \cite{stapp_bells_1975}. The reason is that Bell's
theorem clearly shows that far away measurements indeed affect the
outcomes of a nearby measurement for entangled systems. In other words,
quantum mechanics is non-local, which seems to be incompatible with
the principles behind relativity \cite{maudlin_quantum_2011}. Furthermore,
both quantum mechanics and relativity are tremendously successful
theories, from an empirical point of view. 

As we saw, Bell's setup differs significantly from the double-slit
experiment. First, it does not allow for direct influences between
the observable quantities, which if allowed would be in direct conflict
with special relativity. Instead, the absence of a joint probability
distribution (and therefore of a non-contextual hidden-variable theory)
comes from the non-trivial correlations imposed by QM (and experimental
observations). But, more importantly, Bell's setup provides a case
where a system with two (or more) parts can have such non-trivial
correlations even when the measurements of those parts are spacelike
separated. 

We end this section with one last example of contextuality, the Kochen-Specker
paradigm \cite{kochen_problem_1967}. In quantum mechanics, projection
operators\footnote{A Hermitian operator $P$ is a projection operator if $P^{2}=P$.}
constitute the ``simplest'' type of measurement possible. For example,
for a spin $1/2$ particle, its Hilbert space is two dimensional.
In this Hilbert space, the projector $P_{z}=|+\rangle\langle+|$
is an observable with eigenvalues $0$ and $1$, corresponding to not
having and having the property ``spin $+$ in direction $\hat{\mathbf{z}}$.''
So, projection operators correspond to measurements whose outcomes
tell you whether the system has the property measured ($0$) or not
($1$). 

In their paper, Kochen and Specker\index{Kochen-Specker theorem}
asked whether it is possible to assign values $0$ or $1$ to projection
operators in a way that is consistent. To show that this is not possible,
they used a Hilbert space of dimension three and a total of $117$
projectors. However, to understand how their results come about, we
show here a simpler version of $18$ projectors in a four dimensional
Hilbert space, due to Cabello,
Estebaranz, and Alcaine \cite{cabello_bell-kochen-specker_1996}. 
In this version, we have the set of projection
operators $P_{i}$ with corresponding dicotomic random variables $\mathbf{V}_{i}$
taking values $0$ or $1$ depending on whether the property is false
or true\footnote{Technically, it is not necessary to use random variables, since the Kochen-Specker example uses deterministic (probability one) events. The use of random variables extends this setup to more realistic situations, where probability one events are never observed \cite{larsson_kochen-specker_2002,de_barros_inequalities_2000}.}. The index $i$ in $P_{i}$ corresponds to a vector in the
four dimensional space where $P_{i}$ projects onto. Below is a list
of combinations of random variables, and we see that each line corresponds
to a set of projectors that commute, and can therefore be measured
simultaneously, though this is not true for projectors in different
lines.

\begin{align*}
\mathbf{V}_{0,0,0,1}+\mathbf{V}_{0,0,1,0}+\mathbf{V}_{1,1,0,0}+\mathbf{V}_{1,-1,0,0} & =1,\\
\mathbf{V}_{0,0,0,1}+\mathbf{V}_{0,1,0,0}+\mathbf{V}_{1,0,1,0}+\mathbf{V}_{1,0,-1,0} & =1,\\
\mathbf{V}_{1,-1,1,-1}+\mathbf{V}_{1,-1,-1,1}+\mathbf{V}_{1,1,0,0}+\mathbf{V}_{0,0,1,1} & =1,\\
\mathbf{V}_{1,-1,1,-1}+\mathbf{V}_{1,1,1,1}+\mathbf{V}_{1,0,-1,0}+\mathbf{V}_{0,1,0,-1} & =1,\\
\mathbf{V}_{0,0,1,0}+\mathbf{V}_{0,1,0,0}+\mathbf{V}_{1,0,0,1}+\mathbf{V}_{1,0,0,-1} & =1,\\
\mathbf{V}_{1,-1,-1,1}+\mathbf{V}_{1,1,1,1}+\mathbf{V}_{1,0,0,-1}+\mathbf{V}_{0,1,-1,0} & =1,\\
\mathbf{V}_{1,1,-1,1}+\mathbf{V}_{1,1,1,-1}+\mathbf{V}_{1,-1,0,0}+\mathbf{V}_{0,0,1,1} & =1,\\
\mathbf{V}_{1,1,-1,1}+\mathbf{V}_{-1,1,1,1}+\mathbf{V}_{1,0,1,0}+\mathbf{V}_{0,1,0,-1} & =1,\\
\mathbf{V}_{1,1,1,-1}+\mathbf{V}_{-1,1,1,1}+\mathbf{V}_{1,0,0,1}+\mathbf{V}_{0,1,-1,0} & =1.
\end{align*}
Now, since $\mathbf{V}_{i}$ is either $0$ or $1$, we can sum all
the random variables on the left hand side of the previous set of
equations, and because each random variable appears exactly twice,
this must be an even number. However, because we have only $9$ equations,
the sum of the right hand side yields an odd number. This is clearly
a contradiction, and as with the spin case at the beginning, the problem
comes from identifying a random variable (say, $\mathbf{V}_{0,0,0,1}$)
in a given experimental context (say, when measured together with
$\mathbf{V}_{0,0,1,0}$, $\mathbf{V}_{1,1,0,0}$, $\mathbf{V}_{1,-1,0,0}$,
as in line one) with the same variable in a different experimental
context (in our example, $\mathbf{V}_{0,1,0,0}$, $\mathbf{V}_{1,0,1,0}$,
$\mathbf{V}_{1,0,-1,0}$, as in line two). In other words, the Kochen-Specker
theorem shows that the algebra of observables in QM is such that it
is impossible to assign non-contextual values to certain properties
of a system, independent of what the system is. 

To summarize, in this section we presented several different examples of contextual
quantum systems, per our definition of contextuality. We see that
each example presents its own subtle issues. In the double-slit experiment,
contextuality comes mainly from direct influences on the detection
random variable due to a measurement that leads to which-path information\footnote{Though direct influences may not account for the totality of contextual
effects. See \cite{bacciagaluppi_leggett-garg_2014} for an example
in the context of the Leggett-Garg experiment, which is formally related
to the double-slit experiment. }. The direct influence arises from the choice of successively measuring
incompatible observables, $D$ and $P$. In one case, observable $D$
is measured first (i.e., without $P$), and the wave function arriving
to it leads to interference. In the other case, $P$ is measured first,
collapsing the wave function, and leading to a different quantum state
reaching $D$. So, $P$ directly influences $D$. To distinguish it
from the other cases, we call this \emph{contextuality by direct influences}. 

In the Bell-EPR experiment, direct influences are forbidden by special
relativity, and are not observed. Because two observables, say $A$
and $B'$, commute, the changes to the wave function made by $A$
do not affect the outcomes of $B'$. However, the \emph{correlations}
between $A$ and $B'$ are affected by their simultaneous measurement.
So, the most striking feature of Bell's setup was the contextual dependency
of outcomes of experiments in systems that are \emph{spacelike} separated,
suggesting some type of superluminal influence \cite{maudlin_quantum_2011},
or, as Einstein called it, ``\emph{spukhafte Fernwirkung}'' (spooky
action at a distance). The contextuality for observables that
are spacelike separated are called  \emph{nonlocal contextuality, }or simply
\emph{nonlocality}\index{nonlocality}. 

Finally, the Kochen-Specker theorem showed that the algebra of observables
is such that the random variables representing the outcomes of any
state of a measurable experimental system will present contextuality.
This is a fact that comes simply from the observables themselves and,
contrary to the Bell-EPR experiment, which requires an entangled state,
has nothing to do with the state of the system being measured (i.e.,
is state independent). Consistently with the physics literature, we
refer to this as \emph{state-independent contextuality }or simply
\emph{contextuality}. 

We emphasize  that all the physical systems discussed here satisfy
the criteria of contextuality put forth in the Introduction. From
a \emph{device independent} framework \cite{scarani_v_device-independent_2013,brunner_bell_2014},
where experiments involve black boxes with local inputs (settings
for a measurement) and outputs (measurement outcomes), we look only
at random variables. Then, in this framework, the two types of contextuality
would be contextuality by direct influences and contextuality not
by direct influences. So, the double-slit experiment would fall into
the category of contextuality by direct influences, whereas both Bell-EPR
and Kochen-Specker would be contextual not by direct influences. The
distinction between Bell-EPR and Kochen-Specker (i.e., local vs. nonlocal,
state-dependent vs. state-independent) is based on physical principles,
and not on probabilistic principles alone. 

We saw that exemplary physical systems exhibit contextuality, but
from a physical point of view their contextuality is different. This
leads to the definition of at least three types of contextuality:
contextuality by direct influences, non-locality, and state-independent
contextuality. In the next section we examine quantum cognition, and
discuss how quantum contextuality shows up in them.

\section{Contextuality in quantum cognition\label{sec:Contextuality-in-QC}}

In this section we will examine some examples from the burgeoning
field of quantum cognition. Quantum cognition is the use of the mathematical
formalism of quantum mechanics to model cognitive processes. As such,
it should not be confused with the idea that in order to describe
cognition (or consciousness) we need to use quantum mechanics, as
is espoused by Penrose and Hammerof \cite{penrose_emperors_1989,hameroff_orchestrated_1996,hameroff_quantum_1998},
or by Stapp \cite{stapp_mind_2004,stapp_mind_2009,stapp_mind_2014,de_barros_model_2014},
among others. To researchers in quantum cognition, the quantum comes
from the use of the contextual probability theory given by the Hilbert
space formalism to describe cognitive systems, but the underlying
processes that govern it can be classical \cite{khrennikov_quantum-like_2006,khrennikov_quantum-like_2011,de_barros_quantum-like_2012}. 

This section is not intended to be an exhaustive review of this field,
but instead to provide an example to illustrate some of the main features
of quantum cognition models. Quantum models were used to model the
conjunction and disjunction paradoxes \cite{aerts_quantum_2007,pothos_quantum_2009,franco_conjunction_2009,khrennikov_quantum-like_2009},
the Ellsberg paradox \cite{khrennikov_quantum_2009,aerts_quantum_2011,aerts_quantum_2012},
order effects \cite{trueblood_quantum_2011,atmanspacher_order_2012,wang_quantum_2013},
similarity effects \cite{yearsley_towards_2014}, and the Guppy effect
\cite{aerts_quantum_2009,aerts_guppy_2012}, to name a few. The interested
reader is referred to many of the useful reviews on the subject, such
as Khrennikov \cite{khrennikov_ubiquitous_2010}, Busemeyer and Bruza
\cite{busemeyer_quantum_2012}, Khrennikov and Haven \cite{haven_quantum_2013},
and Ashtiani and Azgomi \cite{ashtiani_survey_2015}.

We focus on one of the first applications of the quantum formalism,
the modeling of the violation of Savage's Sure-Thing Principle (STP).
Simply put, Savage's STP states that if a person holds the subjective
view that $A$ is preferred over  $\neg A$ if $B$ is true\footnote{The notation $\neg A$ means ``not $A$.''}, but is
also preferred if $B$ is false, then $A$ should be preferred regardless
of whether the person knows which is true, $B$ or $\neg B$. In Savage's
own words \cite[pg. 21]{savage_foundations_1972}, ``A businessman
contemplates buying a certain piece of property. He considers the
outcome of the next presidential election relevant to the attractiveness
of the purchase. So, to clarify the matter for himself, he asks whether
he should buy if he knew that the Republican candidate were going
to win, and decides that he would do so. Similarly, he considers whether
he would buy if he knew that the Democratic candidate were going to
win, and again finds that he would do so. Seeing that he would buy
in either event, he decides that he should buy, even though he does
not know which event obtains, or will obtain, as we would ordinarily
say. It is all too seldom that a decision can be arrived at on the
basis of the principle used by this businessman, but, except possibly
for the assumption of simple ordering, I know of no other extralogical
principle governing decisions that finds such ready acceptance.''
Formally, let us imagine a $\pm1$-valued random variable \textbf{$\mathbf{A}$
}($\mathbf{A}=1$ is buy and $\mathbf{A}=-1$ is not buy), and let
another $\pm1$-valued variable $\mathbf{B}$ be $1$ if a Democrat
wins $-1$ if a Republican wins.\textbf{ }If \textbf{$\mathbf{A}=1$
}is preferred over $\mathbf{A}=-1$ when\textbf{ $\mathbf{B}=1$}
and also when $\mathbf{B}=-1$, then $\mathbf{A}=1$ is always preferred.

Even though most people would agree with Savage that his principle
``finds such ready acceptance,'' in some experimental conditions,
human decision makers violate it. For example, in Tversky and Shafir's
1992 paper \cite{tversky_disjunction_1992}, participants were told about a
two-step gamble. In the first step, which was compulsory for all players,
there was a 50/50 probability of winning \$200 or loosing \$100. The
second step was not compulsory, and the person could choose whether
or not to make a second gamble with the same odds and payoffs. After
winning the first bet, 69\% of participants chose to place a second
gamble, and after loosing 59\% also chose to gamble a second time.
In terms of probabilities, we have 
\[
P(\mbox{``gamble again''}|\mbox{``won''})=0.69>P(\mbox{``not gamble again''}|\mbox{``won''})=0.31,
\]
\[
P(\mbox{``gamble again''}|\mbox{``lost''})=0.59>P(\mbox{``not gamble again''}|\mbox{``lost''})=0.41.
\]
Since ``gamble again'' is preferred over ``not gamble again''
for both situations, ``won'' or ``lost'' the first step, STP tells
us that ``gamble again'' should be preferred over ``not gamble
again.'' However, in a later time, the same participants were asked
about the second gamble, but this time they were not told whether
they won or lost the first step. Under this unknown condition, 64\%
of the participants rejected the second gamble. But this corresponds to 
\[
P(\mbox{``gamble again''})=0.36<P(\mbox{``not gamble again''})=0.64,
\]
a clear violation of the STP.

Violations of STP are violations of the standard calculus of probability. To see this, imagine you have two sets, $A$ and $B$, and
we define the conditional probability of $A$ given $B$ as 
\[
P\left(A|B\right)=\frac{P\left(A\cap B\right)}{P\left(B\right)},
\]
for $P\left(B\right)\neq0$. In this notation, the STP conditions
are equivalent to 
\begin{equation}
P\left(A\cap B\right)>P\left(\overline{A}\cap B\right),\,\,\,\,\,\,P\left(A\cap\overline{B}\right)>P\left(\overline{A}\cap\overline{B}\right),\label{eq:STP-sets}
\end{equation}
where $\overline{A}$ denotes the complement of $A$. But from the
calculus of probability, 
\[
P\left(A\cap B\right)+P\left(A\cap\overline{B}\right)=P\left(A\right)
\]
and 
\[
P\left(\overline{A}\cap B\right)+P\left(\overline{A}\cap\overline{B}\right)=P\left(\overline{A}\right),
\]
and adding each term in (\ref{eq:STP-sets}) we obtain
\[
P\left(A\right)>P\left(\overline{A}\right).
\]
Notice that the violation of the law of probabilities happens because
we assume there is a joint probability distribution for $A$ and $B$,
which of course in this case is clear there should be, since we are
dealing with the conditional probability of $A$ given $B$ as something
given subjectively by the decision maker. 

How can we model such violations of classical probability theory with
a quantum formalism? The answer to this particular case is given by
the quantum description of the double-slit experiment \cite{khrennikov_quantum-like_2009,aerts_quantum_2009,khrennikov_quantum_2009}
in the MZI paradigm. In the MZI, let us have the statement ``detector
$D_{1}$ is preferred over $D_{2}$'' as corresponding to a higher
probability of detecting a particle in $D_{1}$ instead of $D_{2}$.
In the notation of Section \ref{sec:Contextuality-in-QM}, this corresponds
to 
\begin{equation}
p\left(\mathbf{D}=1\right)>p\left(\mathbf{D}=-1\right).\label{eq:STP-MZI}
\end{equation}
The which-path information is given by $\mathbf{P}$, and for the
MZI we have 
\begin{eqnarray*}
p\left(\mathbf{D}=1|\mathbf{P}=1\right) & = & p\left(\mathbf{D}=-1|\mathbf{P}=1\right)\\
 & = & p\left(\mathbf{D}=1|\mathbf{P}=-1\right)\\
 & = & p\left(\mathbf{D}=-1|\mathbf{P}=-1\right)\\
 & = & \frac{1}{2}.
\end{eqnarray*}
Similarly to violations of STP, because of the symmetry of the probabilities\footnote{This symmetry is not necessary, and we put it here to make the argument
simpler and to have a direct connection to the experimental setup
shown in Section \ref{sec:Contextuality-in-QM}. The MZI can be modified
to introduce biases that would make the probabilistic system non-symmetric,
but would still lead to a violation of STP (using the same mapping
as we have here). }, we have no reason to prefer $\mathbf{D}=1$ over $\mathbf{D}=-1$
or \emph{vice versa}, as from the symmetry and classical probability
theory we have that $p\left(\mathbf{D}=1\right)=p\left(\mathbf{D}=-1\right)$,
in disagreement with (\ref{eq:STP-MZI}). So, to model a violation
of STP for human decision makers, all we need to do is map the MZI,
with responses ``gamble again'' or ``not gamble again'' corresponding
to $\mathbf{D}=1$ and $\mathbf{D}=-1$, and which path information
corresponding to ``won'' or ``lost.'' The ``won'' and ``lost''
intermediate states are though as mental states that ``collapse''
once the decision maker becomes knowledgeable of the outcome of the
first gamble. 

We remark that, as far as we know, all the quantum cognition models
have similar characteristics to the MZI. What we mean by this is the
following. In the MZI the violation of classical probability theory
comes from two incompatible sets of experimental data due to different contexts. In one context, which-path
information is not present, and the quantum state reaching detectors
$D_{1}$ and $D_{2}$ is in a superposition with components from both
paths. In the other context, a which-path measurement is performed,
and the wave function collapses, with a corresponding loss of quantum
superposition. So, differences in probabilities usually come from
collapse/no-collapse of the wave function due to a measurement. 

Some researchers have suggested that other quantum-like effects may
exist in cognitive systems, such as entanglement \cite{aerts_quantum_2009,bruza_is_2009,aerts_quantum_2013}.
Entanglement comes from states such as (\ref{eq:psi-EPR}), where
for $N$-partite systems ($N\geq2$), the outcomes of a property of
a subsystem are connected to another property of another subsystem
in a way that cannot be explained by common causes (i.e., hidden variables),
as it was the case with the Bell-EPR setup. We have argued elsewhere
that, because we cannot rule out other physical mechanisms, such types
of entanglement are not as unexpected as the quantum mechanical ones
\cite{de_barros_quantum_2009}, and in fact can be derived by classical-like
models \cite{de_barros_quantum-like_2012}. Exactly because of this
reason, there are no principles denying violations of the no-signaling
condition (corresponding to direct influences between different subsystems),
as we have in actual quantum systems \cite{oas_exploring_2014}. Furthermore,
it seems that most of the cases of violations of inequalities such
as (\ref{eq:CHSH}) also violate a form of the no-signaling condition,
and no contextuality from entanglement is detectable, suggesting that
direct influences are more important in quantum cognition \cite{dzhafarov_is_2015}. 

We can also emphasize that, in quantum cognition, violations of the CHSH  inequalities do not necessarily mean nonlocality\index{nonlocality}. To demonstrate non-locality, one needs to not only show violations of the CHSH of events that are spacelike separated (a seemingly impossible task for cognitive events), but also that such violations are not subject to standard loopholes  \cite{larsson_loopholes_2014}. To see how difficult this task is, attempts to create a loophole free test of nonlocality for quantum systems have yet to be successful, even after many decades of intensive research \cite{stobinska_towards_2014}. One can only imagine the technical and conceptual difficulties that would make it hopeless to show nonlocality for cognitive systems. Other difficulties are also present in the Kochen-Specker system \cite{kirchmair_state-independent_2009}. So, from an empirical point of view, it seems that the predominant ``quantum'' effect in cognition is related to the MZI. 

Let us end this Section with one important example. To motivate it,
let us recall that quantum cognition relies on using the quantum mechanical
mathematical apparatus to social systems. However, this seems too
constraining, leaving out many situations that would not be describable
by the formalism. As an example, mentioned in the previous paragraph,
there are no reasons to require cognitive systems to satisfy the no-signaling
condition, and we even have evidence that it is violated for some
cognitive systems. However, the no-signaling condition is a direct
consequence of the quantum formalism: we can derive it from the structure
of the Hilbert space. Furthermore, other forms of ``superluminal''
communications are strictly forbidden by quantum mechanics. This is
the case with the no-cloning theorem \cite{dieks_communication_1982};
if cloning were possible, one could devise a method of sending communications
between Alice and Bob in the EPR setup discussed above. But we have
no \emph{a priori} reason to rule out state cloning for social systems.
So, is the quantum mechanical apparatus too constraining?

As a toy example, we refer back to the $\pm1$-valued random variables,
$\mathbf{X}$, $\mathbf{Y}$, and $\mathbf{Z}$, discussed in the
Introduction. It is possible to concoct artificial (but reasonable)
cases where those random variables have no joint probability distribution,
presenting contextuality \cite{de_barros_decision_2014}. Furthermore,
it is also possible to show that, under reasonable assumptions, neural
models that lead to similar outcomes described by quantum mechanics
\cite{de_barros_quantum-like_2012,de_barros_response_2014,suppes_phase-oscillator_2012}, may also
generate correlated variables $\mathbf{X}$, $\mathbf{Y}$, and $\mathbf{Z}$
that have no joint \cite{de_barros_joint_2012,de_barros_beyond_2015}.
However, such simple example cannot be described by quantum mechanics
(unless we create a contrived model of it \cite{de_barros_decision_2014}),
since the  existence of observed correlations corresponds to pairwise
commutations between the quantum operators representing $\mathbf{X}$,
$\mathbf{Y}$, and $\mathbf{Z}$, and from the algebra of operators
it follows that all three variables $\mathbf{X}$, $\mathbf{Y}$,
and $\mathbf{Z}$ are simultaneously observable. Since they are all
simultaneously observable, a joint probability distribution must exist.
So, the quantum formalism rules out situations such as those described
in \cite{de_barros_joint_2012,de_barros_decision_2014,de_barros_beyond_2015}\footnote{For a more detailed discussion of the neural model and its connection
to quantum cognition and to the issues mentioned in this paragraph,
the reader is referred to \cite{de_barros_quantum_2015}.}. 

To summarize, we sketched in this section how the formalism of quantum
mechanics is often used in quantum cognition. We claimed that among the
many different cases where contextuality shows up in quantum mechanics,
it seems that the only relevant case may be the double-slit experiment. We also saw that the quantum formalism may present too many restrictions
to certain contextual situations. In the next section, we discuss
an alternative formalism that we have proposed in previous papers:
negative probabilities.

\section{Describing contextual probabilities\label{sec:Describing-QC}}

We saw that contextuality is a key factor in quantum cognition, and
the main push for using the quantum mechanical mathematical apparatus
was the better fit it provided for certain experiments. This should
not come too much as a surprise, as this apparatus was developed to
deal with systems that are contextual, such as the double-slit experiment.
But we also saw that there may be cases where  quantum mechanics imposes
too many restrictions that would make its Hilbert space formalism
inadequate to represent them. So, the question is how to develop a
theory of probabilities that have the same ability to describe contextual
systems as quantum mechanics, but also has the flexibility of describing
the systems discussed above. 

There are many attempts to describe quantum contextual systems, such as contextuality by default\cite{dzhafarov_all-possible-couplings_2013,dzhafarov_contextuality_2014-1,dzhafarov_random_2013} or upper and lower probabilities \cite{de_barros_probabilistic_2010,hartmann_entanglement_2010,holik_discussion_2014,suppes_existence_1991}. Here we present a possible theory, first
appearing in Physics in the works of Wigner \cite{wigner_quantum_1932},
but later on considered more seriously by Dirac \cite{dirac_bakerian_1942}
and Feynman \cite{feynman_negative_1987} (for a historical but not
up-to-date survey of negative probabilities in physics, the reader
is referred to \cite{muckenheim_review_1986}).

As mentioned, negative probabilities showed up in quantum mechanics,
when Wigner asked which joint probability distributions for momentum
and position would result in the same outcomes predicted by quantum
statistical mechanics \cite{wigner_quantum_1932}. When such joint
probability distributions were computed for some physical systems
(see, e.g. \cite{suppes_probability_1961}), it became clear that
they could take negative values, and were therefore discarded as non-physical probabilities
(Wigner called them quasi-probability distributions). Though  agreeing with Wigner's claim
of no physical meaning, Dirac thought negative probabilities
could be as useful as negative numbers were in mathematics, and attempted
to apply them to the description of quantized fields \cite{dirac_bakerian_1942},
with no success. Decades later, Feynman also tried to use negative
probabilities, but, to his disappointment, thought that they did not
offer any new insights or results in quantum-field theory \cite{feynman_negative_1987}.
However, since then, some researchers have been using negative probabilities
to help understand certain physical systems, mainly when violations
of classical probabilities occur because of contextuality \cite{scully_feynmans_1994,spekkens_negativity_2008,hartle_quantum_2008,abramsky_sheaf-theoretic_2011,veitch_negative_2012,al-safi_simulating_2013,halliwell_negative_2013,veitch_negative_2013,zhu_negative_2013,abramsky_operational_2014,oas_exploring_2014,bracken_waiting_2014,loubenets_context-invariant_2015}.

Let us start by defining negative probabilities\index{negative probabilities}
(we follow \cite{de_barros_negative_2015,de_barros_quantum_2015}).
We start with a preliminary definition. 

\begin{defn}[compatibility]
Let $\Omega$ be a finite set,
$\mathcal{F}$ an algebra over $\Omega$, and let $\left(\Omega_{i},\mathcal{F}_{i},p_{i}\right)$,
$i=1,\ldots,n$, a set of $n$ probability spaces, $\mathcal{F}_{i}\subseteq\mathcal{F}$
and $\Omega_{i}\subseteq\Omega$. Then $\left(\Omega,\mathcal{F},p\right)$,where
$p$ is a real-valued function, $p:\mathcal{F}\rightarrow\left[0,1\right]$,
$p\left(\Omega\right)=1$, is \emph{compatible} with the probabilities
$p_{i}$'s iff 
\[
\forall\left(x\in\mathcal{F}_{i}\right)\left(p_{i}\left(x\right)=p\left(x\right)\right).
\]
The marginals $p_{i}$ are called \emph{viable} iff $p$ is a probability
measure. 
\end{defn}

The idea of the previous definition is that for contextual systems,
our observations are always in subspaces of a larger sample space
$\Omega$. If it is not possible to put all the observed marginals
in those systems in a single space, then the marginals are not \emph{viable}\footnote{A term coined in Reference \cite{halliwell_negative_2013}.},
i.e. there does not exists a joint probability distribution over $\Omega$
that explains all correlations.\emph{ }

In QM (and, perhaps, social sciences), the marginals are not always viable. This means that no proper
joint probability distribution exists, but perhaps a real-valued function
(but sometimes negative) $p$ exists that provides all the correct
marginals. This $p$, if normalized, would be a negative probability. 

\begin{defn}[negative probabilities]\index{negative probabilities!definition}
Let $\Omega$ be a finite
set, $\mathcal{F}$ an algebra over $\Omega$, $P$ and $P'$ real-valued
functions, $P:\mathcal{F}\rightarrow\mathbb{R}$, $P':\mathcal{F}\rightarrow\mathbb{R}$,
and let $\left(\Omega_{i},\mathcal{F}_{i},p_{i}\right)$, $i=1,\ldots,n$,
a set of $n$ probability spaces, $\mathcal{F}_{i}\subset\mathcal{F}$
and $\Omega_{i}\subseteq\Omega$. Then $\left(\Omega,\mathcal{F},P\right)$
is a negative probability space, and
$P$ a negative probability, if and only if $\left(\Omega,\mathcal{F},P\right)$
is compatible with the probabilities $p_{i}$'s and
\begin{eqnarray*}
\mbox{N1.} &  & \forall\left(P'\right)\left(\sum_{\omega_{i}\in\Omega}\left|P\left(\left\{ \omega_{i}\right\} \right)\right|\leq\sum_{\omega_{i}\in\Omega}\left|P'\left(\left\{ \omega_{i}\right\} \right)\right|\right)\\
\mbox{N2.} &  & \sum_{\omega_{i}\in\Omega}P\left(\left\{ \omega_{i}\right\} \right)=1\\
\mbox{N3.} &  & P\left(\left\{ \omega_{i},\omega_{j}\right\} \right)=P\left(\left\{ \omega_{i}\right\} \right)+P\left(\left\{ \omega_{j}\right\} \right),~~i\neq j.
\end{eqnarray*}
\end{defn}

In the above definition, the standard axiom of nonnegativity \cite{kolmogorov_foundations_1956}
is replaced with a minimization of the L1 norm of $P$ (we use $P$
for negative joint probability distributions, and $p$ for proper
probability distributions). Intuitively, we minimize the L1 norm to
find a quasi-probability distribution that is as close as possible
to a proper probability distribution, since relaxing the nonnegativity
axiom leads to an infinite number of quasi-probabilities consistent
with the marginals. The value of the \emph{minimum L1 probability
norm} is denoted\emph{ $M^{*}$, }and is mathematically given by $M^{*}=\sum_{\omega_{i}\in\Omega}\left|P\left(\left\{ \omega_{i}\right\} \right)\right|$.
This is a useful quantity, since $P$ is a proper probability (and
therefore $\left(\Omega,\mathcal{F},P\right)$ is a probability space)
if and only if $M^{*}=1$ \cite{de_barros_negative_2015}. Furthermore,
we can think of $M^{*}$ as a measure of contextuality: the larger
its value, the more contextual the system \cite{oas_exploring_2014,de_barros_unifying_2014}. 

Not all contextual systems allow for negative probabilities. For instance,
in references \cite{abramsky_sheaf-theoretic_2011,al-safi_simulating_2013,oas_exploring_2014,loubenets_context-invariant_2015}
it was independently proven that negative probabilities exist if and
only if the marginals $p_{i}$  do not allow for direct influences
(we called such systems contextually biased in \cite{de_barros_quantum_2015}).
An example of a system that allows for direct influences are the MZI
and the double slit. 

So, we see that negative probabilities are a possible extension of
standard probability. It is not clear how this extension can be used
to describe, in general, random variables that are directly influenced
by others. But, in some of the cases treated in quantum cognition, it is possible,
by reasoning in a counterfactual way as to preserve the possibility
of identifying random variables in different contexts (see \cite{de_barros_negative_2014,de_barros_negative_2015}
for a detailed analysis of the MZI with negative probabilities). But
the question remains as to whether negative probabilities provide
any advantages over other approaches. 

Before we continue our exposition of negative probabilities, we should
address the issue of interpretation\index{negative probabilities!interpretations of}, which surely is being asked by
many at this point. There are many different ways to interpret negative
probabilities, such as Khrennikov's quasi-stochastic $p$-adic processes \cite{khrennikov_p-adic_1993,khrennikov_p-adic_1993-1,khrennikov_interpretations_2009},
Abramsky and Brandembuerger's negative and positive types \cite{abramsky_operational_2014},
or Szekely's square-root of a coin \cite{szekely_half_2005}, to mention a few.
Here we take a more subjective (and pragmatic) approach, where negative
probabilities are seem as a computational device to help establishing
truth values to propositions (say, the proposition ``the random variable
$\mathbf{A}$ has value $1$''). As such, the minimization of the
L1 norm is nothing but a requirement that this computation should
give us a quasi-probability distribution that is as close as possible
to a (non-existent) proper probability distribution. In other words,
as Feynman and Dirac, we see negative probabilities as a computational
device, without necessarily having a meaning. 

We now turn to an example, first analyzed in \cite{de_barros_decision_2014},
and based on the three $\pm1$-valued random variables $\mathbf{X}$,
$\mathbf{Y}$, and $\mathbf{Z}$. Imagine a decision maker, here named
Deanna, who  wants to invest in the stocks of three companies,
$X$, $Y$, and $Z$. Knowing nothing about the stock market, Deanna
hires three experts, Alice, Bob, and Carlos, to give her advice. Their
range of expertise overlaps, but are not the same: Alice knows only
about $X$ and $Y$, but knows nothing about $Z$; Bob knows about
$X$ and $Z$, but not about $Y$; and Carlos knows about $Y$ and
$Z$, but not $X$. All experts agree that the chances of $X$, $Y$,
and $Z$ going up are the same as them going down. Alice also tells
Deanna that she thinks that whenever $X$ goes up, $Y$ is sure to
go down, and vice versa. Bob tells her that whenever $X$ goes one
way, $Z$ goes the other way with probability $3/4$, and they go
the same way with probability $1/4$. Finally, Carlos tells her that
he sees no relationship whatsoever between $Y$ and $Z$. Associating
a $+1$ value of a random variable with stock-values going up, and
$-1$ with going down, Deanna has the following expectations for $\mathbf{X}$,
$\mathbf{Y}$, and $\mathbf{Z}$. 
\begin{equation}
E\left(\mathbf{X}\right)=E\left(\mathbf{Y}\right)=E\left(\mathbf{Z}\right)=0,\label{eq:XYZ-expectations}
\end{equation}
\begin{equation}
E\left(\mathbf{XY}\right)=-1,\label{eq:XY-moment}
\end{equation}
\begin{equation}
E\left(\mathbf{XZ}\right)=-\frac{1}{2},\label{eq:XZ-moment}
\end{equation}
and 
\begin{equation}
E\left(\mathbf{YZ}\right)=0.\label{eq:YZ-moment}
\end{equation}
It is not hard to see, using Suppes and Zanotti's inequalities \cite{suppes_when_1981},
that there is no joint probability distribution for $\mathbf{X}$,
$\mathbf{Y}$, and $\mathbf{Z}$ consistent with the expectations
(\ref{eq:XYZ-expectations})--(\ref{eq:YZ-moment}). 

How should Deanna proceed? There are some alternatives in the literature,
but perhaps the most common one would be the Bayesian approach, where
Deanna starts with a prior probability distribution which is updated
with the experts opinions \cite{morris_decision_1974,morris_combining_1977}.
However, as we pointed out elsewhere, this approach has problems.
First, the triple moment $E\left(\mathbf{XYZ}\right)$ is invariant
under Bayesian updates of the pairwise moments. This means that whatever
values of triple moments Deanna starts with, those values are not
updated \cite{de_barros_decision_2014}. This lack of update presents
problems when we expect Deanna to get information about the triple
moment in cases where the three experts agree. For instance, if Alice,
Bob, and Carlos all say that the stocks are perfectly correlated,
Deanna's update should lead to $E\left(\mathbf{XYZ}\right)=1$. Furthermore,
for certain ``inconsistent'' (with a joint) correlations given by
Alice, Bob, and Carlos, weakening them would lead to a restricted
value of the triple moment \cite{de_barros_quantum_2015}. But how
to extract such information from the inconsistent beliefs given by
the experts? When a joint negative probability distribution is constructed,
not any distribution is allowed, but only those minimizing the L1
norm\index{negative probabilities!L1 norm}. We can show that, for the above example, minimizing the L1 norm
constrains the values of the triple moment to be in the range 
\[
-\frac{1}{4}\leq E\left(\mathbf{XYZ}\right)\leq\frac{1}{2}.
\]
Therefore, negative probabilities provide information about the range
of values of the triple moment that is not part of the standard Bayesian
update. With such range, it should be possible, under certain conditions,
to formulate a Dutch book.

\section{Final Remarks\label{sec:Final-Remarks}}

In this paper we examined different examples of contextuality in physics,
namely the double-slit experiment, in the form of the Mach-Zehnder
interferometer (MZI), the Bell-EPR entanglement experiment, and the
Kochen-Specker theorem. We argued that among those cases, the one
that relates more closely to what is usually done in quantum cognition
is the MZI, and that both the Bell-EPR and the Kochen-Specker paradigms
have only marginal interest (from an empirical point of view). 

We also discussed the restrictions on types of systems that can be
modeled by the Hilbert space formalism of quantum mechanics. Such
formalism cannot model certain systems that are not, in principle,
forbidden by any cognitive or behavioral principle. It also implies
constraints such as the no-cloning theorem \cite{dieks_communication_1982},
the no-signaling condition \cite{redhead_relativity_1986}, and the
monogamy of quantum correlations (from entanglement) \cite{seevinck_monogamy_2010},
to name a few cases. Such constraints are almost necessary for Physical
systems, or there would be severe conflicts with the (empirically
verified) theory of relativity, but they are not at all necessary
for cognitive systems. In fact, the Hilbert space formalism is so
restrictive that it even forbids the simple three-random-variable
example we showed in Section \ref{sec:Describing-QC}.

With the examples presented, we are not attempting here to discourage
the use of the quantum formalism in cognition. We believe that the
quantum mathematical structure inspired many interesting results in
quantum cognition, and the large volume of papers in the subject attest
to its importance. Our goal is instead to point out that there are
other tools, such as negative probabilities, that should not be neglected,
and perhaps studied side-by-side with the quantum structures (this
is the subject of our other paper in this book \cite{oas_survey_2015}),
and to challenge the quantum interactions community to think about
cases where the quantum formalism may be inadequate or (at least)
cumbersome. For instance, the examples we presented raise some questions
about the quantum models. Can the (apparent) advantage of negative
probabilities, in certain examples, over the Bayesian approach be
reproduced with the quantum formalism? What principles would have
to be added to them? Do human-decision makers follow a process similar
to the minimization of the L1 norm for inconsistent situations? If
so, how would such process be described in the quantum formalism?
Those are open questions that would need to be addressed.

\paragraph{Acknowledgments. }

We benefited tremendously from our collaboration with Janne Kujala and Ehtibar Dzhafarov, and we thank them both, and in particular Professor Dzhafarov for his kind invitation to participate in the 2014 Winer Memorial Lectures. We also thank Jerome Busemeyer, Samsom Abramsky, Louis Narens, Guido Bacciagaluppi,
Andrei Khrennikov, Arkady Plotnitsky, Claudio Carvalhaes, and Patrick Suppes for discussions. 

\bibliographystyle{plain}
\bibliography{Quantum}
%\blankpage
%\printindex[aindx]                 % to print author index
\printindex  

%\bibliographystyle{plain}
%\bibliography{Quantum}

\end{document}